\documentclass[10pt]{article}
\usepackage[cp1250]{inputenc}
\usepackage{amsmath}

\usepackage[dvips]{graphicx}
\usepackage{amssymb}

\begin{document}

\title{A broad-band laser-driven double Fano system -- photoelectron spectra}

\author{K. Doan Quoc$^{1,2}$, V. Cao Long$^{1,*}$ and W. Leo\'nski$^1$}

\date{}

\maketitle

\begin{center}
%
%
{\small\it $^1$Quantum Optics and Engineering Division, Institute of Physics,  University of Zielona G\'ora, ul.~Prof.~A.~Szafrana~4a, 65-516 Zielona G\'ora, Poland\\
$^2$Quang Tri Teacher Training College, Km3, Highway No 9, Dong Ha, Quang Tri, Vietnam}\\

$^*$E-mail: vanlongcao@yahoo.com
\end{center}

\begin{abstract}
Fano profiles, which with their asymmetric character, have many potential applications in technology. The design of Fano profiles into optical systems may create new non-linear and switchable metamaterials, high-quality optical waveguides, ultrasensitive media for chemical or biosensing etc.  In this paper, we consider an external field driven double Fano model, in which, instead of one autoionizing state, there are two discrete states embedded in one continuum. We assume further that the external electromagnetic field can be decomposed into two parts: a deterministic (or coherent) part and a randomly fluctuating chaotic component, which is a $\delta$-correlated, Gaussian, Markov and stationary process (white noise). This assumption corresponds to the case of the real multimode laser operating without any correlation between the modes. We solve a set of coupled stochastic integro-differential equations involving the Fano model with two discrete levels. We derive an exact formula by determining the photoelectron spectrum and compare it with the results obtained earlier in our previous papers.
\end{abstract}

%
%
%
\section{Introduction}
Over the last three decades, we have noticed a particular interest in the research on different ionization processes of atoms in laser fields. One of the most interesting examples is the so-called laser-induced autoionization, on which several papers have already been published, including the study of the detailed characteristics of electron and photon spectra associated with photoexcited autoionization. The Fano model \cite{B1} for a number of discrete states and one continuum that can be diagonalized is the most commonly used atomic model. Fano diagonalization, based on the Coulomb mixing of ionizing states with the continuum, leads to a nontrivial structure of the latter [1-5]. Systems comprising autoionizing levels can behave in a diverse and nontrivial way, leading to interesting physical phenomena, such as quantum interferences discussed in [1-8](and the references cited therein), electromagnetically induced transparency used in light slowdown \cite{B9}, and the quantum anti-Zeno effect \cite{B10}. The main feature of the Fano profile is its asymmetric character. Since its discovery, the Fano profile has been found in several functioned materials as plasmonic nanoparticles, photonic crystals and electromagnetic metamaterials. The unique properties associated with this asymmetric line shape lead to potential applications in a wide range of technologies \cite{B11}. The chemical and biological sensors are the most straightforward examples of this fact. An interesting review on Fano profiles in nanoscale structures is given in \cite{B12}.\\
 
The advantage of a model comprising a Fano-diagonalized atom + laser light is that it can be solved exactly, even if relaxation mechanisms are included in the standard way. Because of the very complicated microscopic nature of the relevant relaxation mechanisms, they are usually modeled by a classical time-dependent random process. After this process, dynamical equations involved become stochastic differential equations. Except in certain special cases, e.g. chaotic white noise \cite{B13} or pre-Gaussian noise \cite{B14}, these equations cannot be solved in finite terms.\\

In \cite{B3}, we treated the laser field as white noise interacting with the Fano autoionization model. Then, a set of coupled stochastic integro-differential equations was solved exactly. Applying the obtained solutions, we determined  the exact photoelectron spectrum. In this work, we extend the formalism introduced in \cite{B3} to the case of a double Fano system \cite{B2}, in which, instead of one autoionizing state, there are two discrete states embedded in one continuum. As it has been emphasized in \cite{B3}, the model of white noise for the field is interesting by itself because it describes the electric field amplitude of the multimode laser operating without correlation between the modes. This white noise model has been also used in our other paper \cite{B15}, where in a $\Lambda$-like configuration of electromagnetically induced transparency, the laser control field has the white noise component. Then the considered model is more realistic experimentally, because the amplitudes of the real laser light always contain some fluctuation component. As in \cite{B3}, we determine in this paper the exact photoelectron spectrum and compare it with the results obtained earlier \cite{B2,B3}.\\

The paper is organized as follows: in the second section, we describe certain aspects of our model and derive a set of equations for atomic operators involved in the problem. These equations are more accurate than those introduced in \cite{B3}. In the third section, our results are presented and discussed. Instead of presenting a complicated final formula, we are concerned with two interesting physical limits, i.e. for an infinite and finite value of the asymmetry parameter introduced by Fano \cite{B1} and generalized for the double Fano system in \cite{B2}. The final section presents our conclusions.

\section{A model for a double Fano system}
We start from the model shown in figure 1, where besides the ground state $|0\rangle$, there are two discrete states $|1\rangle$ and $|2\rangle$ lying above the ionisation threshold and interacting with the continuum by means of configurational Coulomb interaction. The two levels, as well as the continuum, are coupled to the ground state by a strong laser beam of frequency $\omega_L$.

\begin{figure}[!h]
    \begin{center}
                {\includegraphics[scale=0.75]{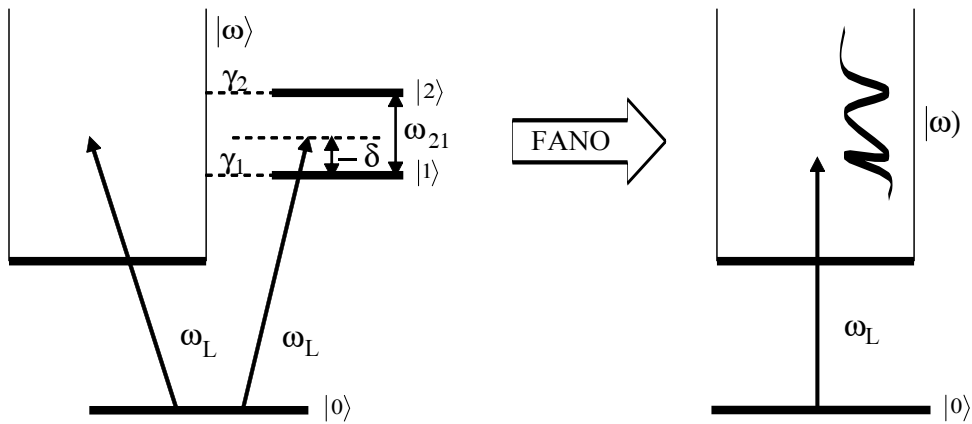}}\label{f1}
    \end{center}
   \caption{Simplified atomic level diagram. Configuration-interaction coupling of levels $|1\rangle$ and $|2\rangle$ to $|\omega\rangle$ leads to the double Fano continuum $|\omega)$. This continuum is coupled to discrete state $|0\rangle$ by a laser of frequency $\omega_L$}
\end{figure}

Following \cite{B2} we can obtain the double Fano profile describing the density of the continuum states. The state of this new structured continuum is denoted by round brackets $|\omega)$ (see the diagram on the right side of figure 1). The matrix element $\Omega(\omega)$ that describes the coupling of the ground state with this new state (the effective Rabi frequency) is given by
\begin{equation}\label{E1}
\Omega(\omega)=\frac{\Omega_0}{\sqrt{4\pi\Gamma}}\Big(\frac{A_+}{\omega-\omega_+}+\frac{A_-}{\omega-\omega_-}+\frac{1}{Q+i}\Big),
\end{equation}
where $\Gamma=\Gamma_1+\Gamma_2$ is the total autoionization rate, the effective asymmetry parameter is expressed by $Q=(q_1\Gamma_1+q_2\Gamma_2)/\Gamma$ and $\omega_\pm$ are the complex roots of the denominator of Eq.(1.1) given by \cite{B2}
\begin{equation}\label{E2}
\omega_\pm=\frac{\omega_1+\omega_2\pm\theta}{2}+i\frac{\Gamma\pm\phi}{2},
\end{equation}

\begin{equation}\label{E3}
\begin{split}
\phi=\frac{1}{\sqrt{2}}\Big\{\big[(\omega_{21}^2-\Gamma^2)^2+4\omega_{21}^2(\Gamma_2-\Gamma_1)^2\big]^{1/2}-\omega_{21}^2+\Gamma^2\Big\}^{1/2},\\
\theta=\frac{1}{\sqrt{2}}\Big\{\big[(\omega_{21}^2-\Gamma^2)^2+4\omega_{21}^2(\Gamma_2-\Gamma_1)^2\big]^{1/2}+\omega_{21}^2-\Gamma^2\Big\}^{1/2},
\end{split}
\end{equation}
where $\omega_{21}=\omega_2-\omega_1$, $\omega_1$ and $\omega_2$ are the bare energies of the discrete atomic states.\\
The complex amplitudes $A_\pm$ are given by the following expressions:

\begin{equation}\label{E4}
A_\pm=\frac{\Gamma}{2}\Big(1\pm\frac{\omega_{21}K+i\Gamma}{\theta+i\phi}\Big),
\end{equation}
where
\begin{equation}\label{E5}
K=\frac{q_2\Gamma_2-q_1\Gamma_1+i(\Gamma_2-\Gamma_1)}{\Gamma(Q+i)}.
\end{equation}

The form equation(\ref{E1}) of the radiative matrix element $\Omega(\omega)$ is a generalization of the corresponding formula of Rza\.zewski and Eberly [4, 5] to the case of two autoionizing levels, both of which are radiatively coupled to the ground state. In fact, $\Omega(\omega)$ is a superposition of two Lorentzians and a flat background. Thus, for the case of the double Fano profile we have an additional Lorentzian as a result of the presence of an additional autoionising state.\\
For the sake of convenience, we shall denote the new state $|\omega)$ by the ket $|\omega\rangle$. As in \cite{B3}, we start with the Hamiltonian, which describes the model with a bound state lying below the edge of the continuum, the continuum states and these states are coupled together by an external electromagnetic field:
\begin{equation}\label{E6}
H=\hbar\omega_0P_0+\int\hbar\omega C_{\omega\omega}d\omega+\int\Omega(\omega)|0\rangle\langle\omega|d\omega+H.C,
\end{equation}
where $P_0$ and $C_{\omega\omega}$ are occupation-number operators for the ground state and for the continuum states, respectively. The radiative interaction between those states is described by the function $\Omega(\omega)$, which marks how strongly different points of the continuous spectrum are coupled to the bound state, and is given by the formula (\ref{E1}). By $|0\rangle$ we mean the bound state and $|\omega\rangle$ stands for the excited state in the dressed continuum.\\ 
Moreover, we define the following operators 
\begin{equation}\label{E7}
\begin{split}
B_\omega&=|0\rangle\langle\omega|,\\
C_{\omega\omega'}&=|\omega\rangle\langle\omega'|.
\end{split}
\end{equation}
In consequence, the Heisenberg equations of motion for the atomic operators $P_0$, $B_\omega$, $B_\omega^+$, $C_{\omega\omega}$ can be derived. They form a complete set and can be easily found by simple commutations of those operators with Hamiltonian (\ref{E6}). For the sake of simplicity we assume that $\omega_0=0$.\\

In the Heisenberg picture, one obtains linear equations for the dynamical variables, so the equations for the corresponding averaged quantities can be easily found using well-known results, based on the theory of multiplicative stochastic processes. We assume now that $\Omega(\omega)$ has the form:
\begin{equation}\label{E8}
\Omega(\omega)=f(\omega)(E_0+E(t))e^{i\omega_Lt},
\end{equation}
where $E(t)$ is characterized by a Gaussian, Markov and stationary process (white noise),

\begin{equation}\label{E9}
\langle\langle E(t)E^*(t')\rangle\rangle=a\delta(t-t'),
\end{equation}
and $E_0$ is a deterministic coherent component of the laser field. The double brackets in (\ref{E9}) indicate an average over the ensemble of realisations of the process $E(t)$.\\ 

Thus, we consider the following stochastic differential equation
\begin{equation}\label{E10}
\frac{dQ}{dt}=\{A+x(t)B+x^*(t)C\}Q,
\end{equation}
where $Q$ is a vector function of time and $A, B$ and $C$ are constant matrices. As it is known from the multiplicative stochastic process theory, the function $\langle\langle Q\rangle\rangle$ exactly satisfies the nonstochastic equation:
\begin{equation}\label{E11}
\frac{d\langle\langle Q\rangle\rangle}{dt}=[A+a\{B,C\}/2]\langle\langle Q\rangle\rangle,
\end{equation}
where $\{B,C\}$ is the anticommutator of $B$ and $C$. The result contained in equation (\ref{E11}) is well-known in the literature and has been used for example in \cite{B3}.\\
 
For the case discussed here, before performing the averaging procedure, we transform dynamical variables to the rotating frame:
\begin{equation}\label{E12}
\begin{split}
B_\omega&=f^*(\omega)D_\omega e^{-i\omega_Lt},\\
C_{\omega\omega'}&=f(\omega')f^*(\omega)E_{\omega\omega'}.
\end{split}
\end{equation}
The new quantities $D_\omega$, $E_{\omega\omega'}$ together with $P_0$ satisfy the following closed set of equations:

\begin{equation}\label{E13}
\begin{split}
&\frac{dP_0}{dt}=-i\int d\omega|f(\omega)|^2[(E_0+E(t))D_\omega-(E_0^*+E^*(t))D_\omega^+],\\
&\frac{dD_\omega}{dt}=i(\omega_L-\omega)D_\omega-i(E_0^*+E^*(t))P_0+ i\int d\omega'|f(\omega')|^2[(E_0^*+E^*(t))E_{\omega'\omega},\\
&\frac{dE_{\omega\omega'}}{dt}=i(\omega-\omega')E_{\omega\omega'}+i(E_0+E(t))D_{\omega'}-i(E_0^*+E^*(t))D_\omega^+.
\end{split}
\end{equation}
Next, using equation (\ref{E11}) we obtain a system of equations for the stochastic averages of the variables (double brackets have been dropped for convenience):

\begin{equation}\label{E14}
\begin{split}
&\frac{dP_0}{dt}=-aFP_0-ib\int d\omega'|f(\omega')|^2(D_{\omega'}-D_{\omega}^+)+a\int\int d\omega' d\omega"|f(\omega')|^2|f(\omega"))|^2E_{\omega'\omega"},\\
&\frac{dD_\omega}{dt}=-ibP_0+\big[i(\omega_L-\omega)-\frac{aF}{2}\big]D_\omega-\frac{a}{2}\int d\omega'|f(\omega')|^2D_{\omega'}+ib\int d\omega'|f(\omega')|^2E_{\omega'\omega},\\
&\frac{dE_{\omega\omega'}}{dt}=aP_0+ib(D_{\omega'}-D_\omega^+)+i({\omega-\omega'})E_{\omega\omega'}-\frac{a}{2}\int d\omega"|f(\omega")|^2(E_{\omega"\omega'}+E_{\omega\omega"}).
\end{split}
\end{equation}
where $F=\int d\omega|f(\omega)|^2$, $b=|E_0|$. The equation corresponding to the adjoint operator $D_\omega^+$ is easily found from that for $D_\omega$ by the calculation of its complex conjugation. As we compare the formulas (\ref{E9}a) and (\ref{E9}b) in \cite{B3} with our eqns. (\ref{E14}a) and (\ref{E14}b), the later contain additional terms corresponding to the coherent part of the laser field b which were omitted in \cite{B3}.\\

Subsequently, taking the Laplace transform of equation (\ref{E14}) and denoting $|f(\omega)|^2=S(\omega)$, we obtain

\begin{equation}\label{E15}
\begin{split}
&z\tilde{P}_0-1=-aF\tilde{P}_0-ib\int d\omega'S(\omega')(\tilde{D}_{\omega'}-\tilde{D}_{\omega}^+)+a\int\int d\omega' d\omega"S(\omega')S(\omega")\tilde{E}_{\omega'\omega"},\\
&z\tilde{D}_\omega=-ib\tilde{P}_0+\big[i(\omega_L-\omega)-\frac{aF}{2}\big]\tilde{D}_\omega-\frac{a}{2}\int d\omega'S(\omega')\tilde{D}_{\omega'}+ib\int d\omega'S(\omega')\tilde{E}_{\omega'\omega},\\
&z\tilde{E}_{\omega\omega'}=a\tilde{P}_0+ib(\tilde{D}_{\omega'}-\tilde{D}_\omega^+)+i({\omega-\omega'})\tilde{E}_{\omega\omega'}-\frac{a}{2}\int d\omega"S(\omega")(\tilde{E}_{\omega"\omega'}+\tilde{E}_{\omega\omega"}).
\end{split}
\end{equation}
where we introduced the appropriate initial conditions for the probabilities $P_0|_{t=0}=1$, $D_\omega|_{t=0}=E_{\omega\omega'}|_{t=0}=0$. As we introduce a new variable (for convenience we drop also the tilde):
\begin{equation}\label{E16}
A^-(z)=-ibP_0-\frac{a}{2}\int d\omega'S(\omega')D_{\omega'},
\end{equation}
we can write the resulting formula for $D_\omega$:
\begin{equation}\label{E17}
D_\omega=\frac{A^-(z)}{z+ac^*/8-i(\omega_L-\omega)}+\frac{ib}{z+ac^*/8-i(\omega_L-\omega)}\int d\omega'S(\omega')E_{\omega'\omega},
\end{equation}
where
\begin{equation}\label{E18}
\begin{split}
&c=\frac{1}{\Gamma}(iB_++iB_-),\\
&B_\pm=2A_\pm\Big(\frac{A_\pm^*}{i(\Gamma\pm\phi)}+\frac{A_\pm^*}{i\Gamma\pm\theta}+\frac{1}{Q-i}\Big),
\end{split}
\end{equation}
and the equation for $D_\omega^+$ can be expressed in a similar way:
\begin{equation}\label{E19}
D_\omega^+=\frac{A^+(z)}{z+ac/8+i(\omega_L-\omega)}-\frac{ib}{z+ac/8+i(\omega_L-\omega)}\int d\omega'S(\omega')E_{\omega\omega'},
\end{equation}
where $A^+(z)$ is simply a Hermitian conjugate of (\ref{E16}). \\

To solve the whole set of our equations, we assume the separation property for $E_{\omega\omega'}$, i.e.:
\begin{equation}\label{E20}
[z-i(\omega-\omega')]E_{\omega\omega'}=\zeta_\omega(z)+\eta_{\omega'}(z).
\end{equation}
This decomposition is of the greatest importance in the entire procedure of solving the equations in question. For the case of the double Fano profile, the functions $\zeta_\omega$ and $\eta_\omega$ satisfy the following equations, where we introduce $a_0=a/8$:
\begin{equation}\label{E21}
H^+(z,\omega)\zeta_\omega=4a_0P_0-\frac{ibA^+(z)}{z+a_0c+i(\omega_L-\omega)}-\bigg[4a_0+\frac{b^2}{z+a_0c+i(\omega_L-\omega)}\bigg]\int\frac{S(\omega')\eta_{\omega'}}{z+i(\omega'-\omega)}d\omega'
\end{equation}
and
\begin{equation}\label{E22}
H^-(z,\omega)\eta_\omega=4a_0P_0+\frac{ibA^-(z)}{z+a_0c^*-i(\omega_L-\omega)}-\bigg[4a_0+\frac{b^2}{z+a_0c^*-i(\omega_L-\omega)}\bigg]\int\frac{S(\omega')\zeta_{\omega'}}{z-i(\omega'-\omega)}d\omega',
\end{equation}
where
\begin{equation}\label{E23}
\begin{split}
&H^+(z,\omega)=1+a_0d+\frac{b^2d}{4[z+a_0c+i(\omega_L-\omega)]},\\
&H^-(z,\omega)=1+a_0d^*+\frac{b^2d^*}{4[z+a_0c^*-i(\omega_L-\omega)]},
\end{split}
\end{equation}
with
\begin{equation}\label{E24}
d=\frac{1}{\Gamma}\Big(\frac{B_+}{\omega_+-\omega-iz}+\frac{B_-}{\omega_--\omega-iz}+\frac{1}{Q^2+1}\Big).
\end{equation}
From the analytic property of $E_{\omega\omega'}$ we assume that $\zeta_\omega$ must be of the following form
\begin{equation}\label{E25}
\zeta_\omega=\frac{4a_0P_0}{H^+}-\frac{ibA^+(z)}{[z+a_0c+i(\omega_L-\omega)]H^+}+\frac{D^+\big[a_0d+\frac{b^2d}{4[z+a_0c+i(\omega_L-\omega)]}\big]}{H^+}.
\end{equation}
The quantity $\eta_\omega$ satisfies a similar equation for the complex amplitude $D^-$. By insertion the expressions for $\zeta_\omega$, $\eta_\omega$ into equations (\ref{E21}) and (\ref{E22}) we obtain after simple algebra

\begin{equation}\label{E26}
D^+=-\frac{4a_0P_0+ibgA^-(z)+D^-[H^-(z,-i\omega)-1]}{H^-(z,-i\omega)},
\end{equation}
where
\begin{equation}\label{E27}
g=\frac{1}{z+a_0c+i(\omega_+-\omega_L)}+\frac{1}{z+a_0c+i(\omega_--\omega_L)}.
\end{equation}
Of course, $D^-$ satisfies an analogous equation after the substitution $A^-\to A^+$, $H^-\to H^+$. Using (\ref{E16})-(\ref{E26}) and the definitions of $A^+$ we have the following equations 

\begin{equation}\label{E28}
\begin{split}
&(1+a_0e)A^-(z)=-ibP_0+\frac{ia_0bef}{4}(D^++D^-),\\
&(1+a_0e^*)A^+(z)=ibP_0-\frac{ia_0be^*f^*}{4}(D^++D^-),
\end{split}
\end{equation}
where
\begin{equation}\label{E29}
\begin{split}
&e=\frac{1}{\Gamma}\Big(\frac{B_+}{\omega_+-\omega_L-ia_0c^*-iz}+\frac{B_-}{\omega_--\omega_L-ia_0c^*-iz}+\frac{1}{Q^2+1}\Big),\\
&f=\frac{1}{\Gamma}\Big(\frac{B_+}{2\omega_+-iz}+\frac{B_-}{2\omega_--iz}+\frac{1}{Q^2+1}\Big).
\end{split}
\end{equation}
In the similar way for $P_0$ the first equation in (\ref{E15}) gives us
\begin{equation}\label{E30}
P_0=\frac{1-2a_0cP_0}{z}+\frac{ib(e^*A^+(z)-eA^-(z))}{4z}-\frac{D^++D^-}{16z}(b^2ef+b^2e^*f^*+f).
\end{equation}
The results presented above can be applied in a steady state spectrum calculation. Such a spectrum will be discussed in the next section for the case of the strong external field. \\

\section{Photoelectron Spectrum for two Lorentzians}
As it was mentioned before, we discuss the case of the double Fano profile with two Lorentzians . For the model discussed here, we define the long-time photoelectron spectrum as:
\begin{equation}\label{E31}
W(\omega)=\lim_{t\to\infty}C_{\omega\omega}(t).
\end{equation}
The spectrum defined in such a way can be computed directly and can be expressed in a completely analytical form. Thus, equations (\ref{E26}), (\ref{E28}) and (\ref{E30}) are linear algebraic equations and their solutions can be easily found. The spectral distribution of excited electrons is determined from $C_{\omega\omega}(t)$. In the steady state at $t\to\infty$, only the pole at $z=0$  in the Laplace domain contributes. From the definition of $E_{\omega\omega'}$ and its separation property, we obtain: 
\begin{figure}[!h]
    \begin{center}
\includegraphics[scale=0.3]{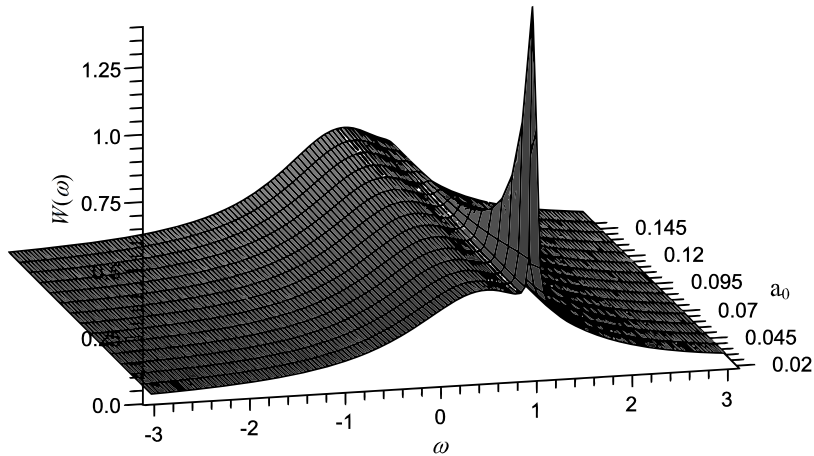}
\label{f2}
   \end{center}
  \caption{Photoelectron spectrum for $\omega_1=\omega_2=0.5$, we assume that $\omega_L=1.0$, autoionisation widths $\Gamma_1=\Gamma_2=0.5$ and the coherent component $b=0.1$.}
\end{figure}

\begin{figure}[!h]
    \begin{center}
\includegraphics[scale=0.3]{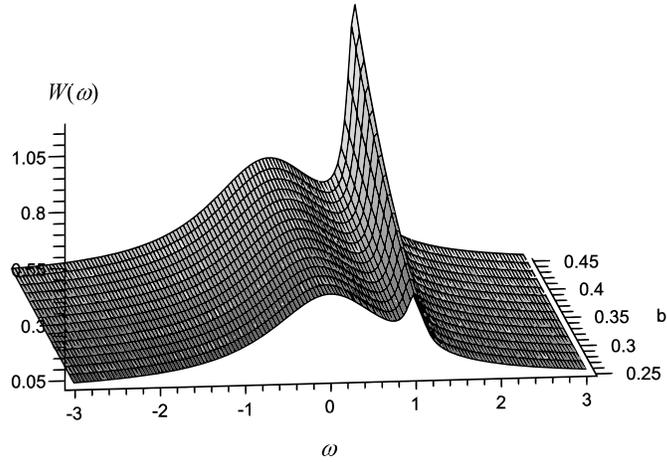}
\label{f3}
   \end{center}
  \caption{Photoelectron spectrum for $\omega_1=\omega_2=0.001$, we assume that $\omega_L=1.0$, autoionisation widths $\Gamma_1=\Gamma_2=0.5$ and the chaotic component $a_0=0.12$.}
\end{figure}
\begin{figure}[!h]
    \begin{center}
\includegraphics[scale=0.3]{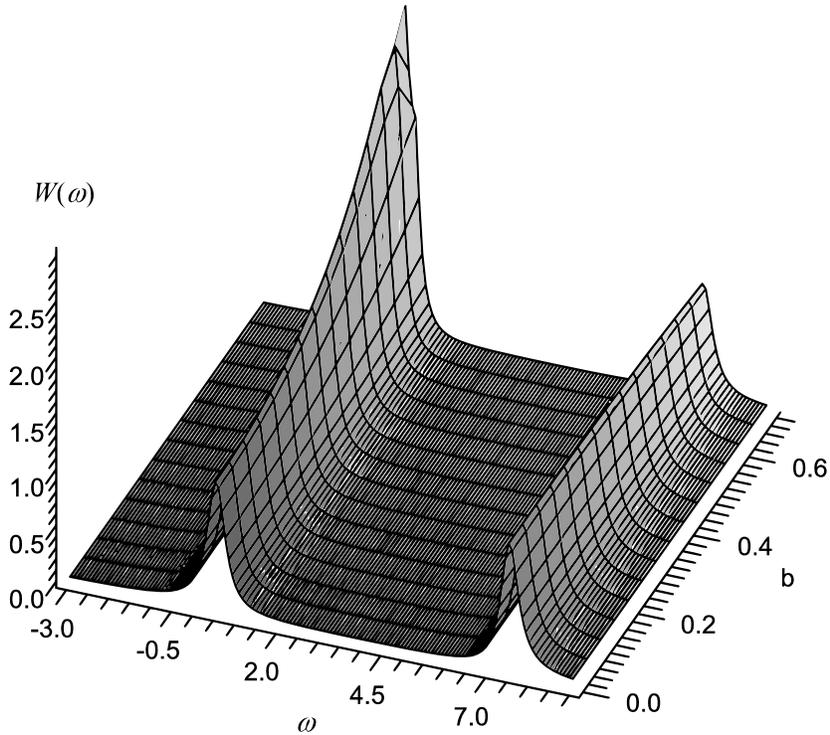}
\label{f4}
   \end{center}
  \caption{Photoelectron spectrum for $\omega_1=0.5$, $\omega_2=7.5$ we assume that $\omega_L=0.05$, autoionisation widths $\Gamma_1=\Gamma_2=0.5$ and the chaotic component $a_0=0.4$.}
\end{figure}
\begin{figure}[!h]
    \begin{center}
\includegraphics[scale=0.35]{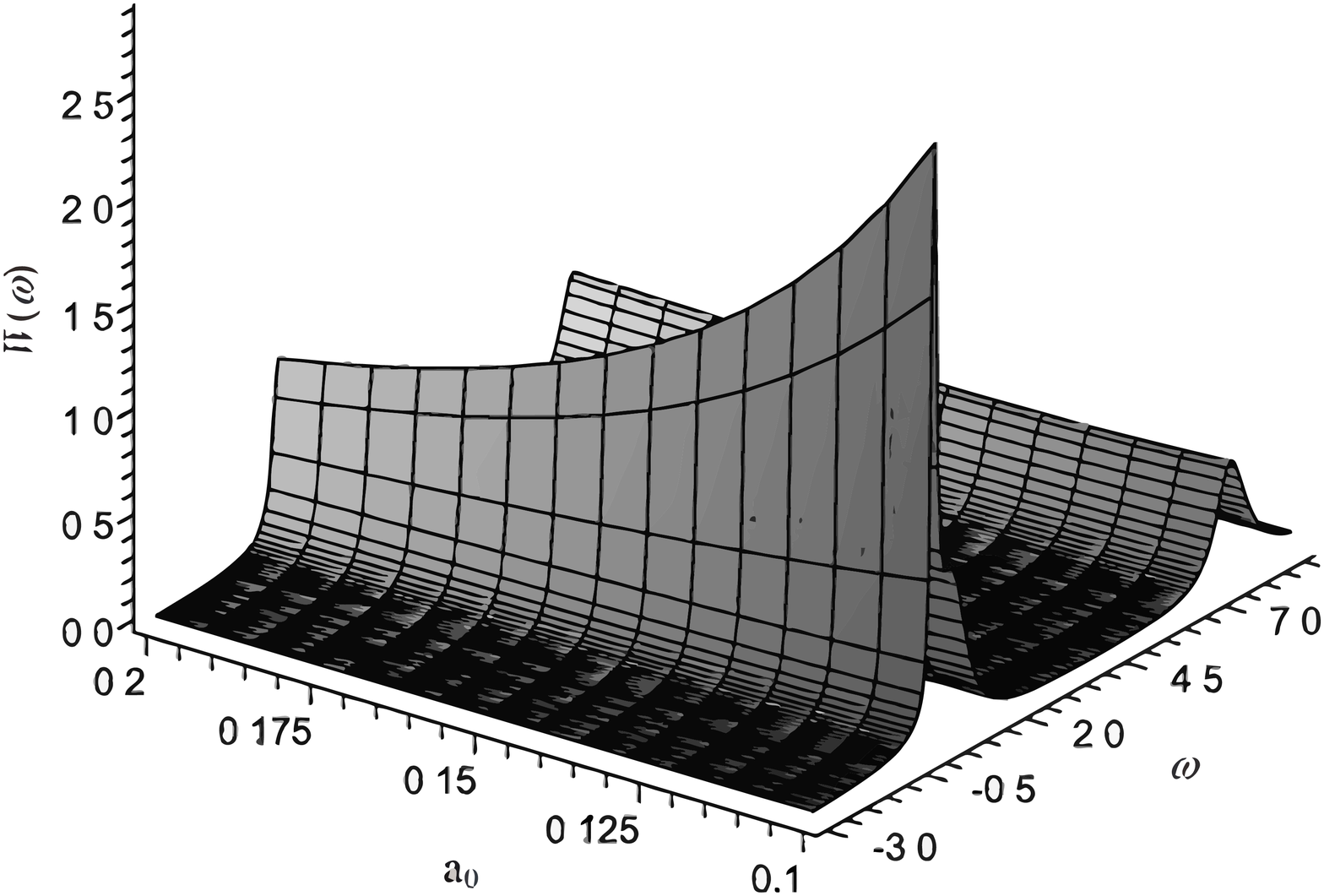}
\label{f5}
   \end{center}
  \caption{Photoelectron spectrum for the case of $\omega_1=0.5$, $\omega_2=7.5$ we assume that $\omega_L=0.05$, autoionisation widths $\Gamma_1=\Gamma_2=0.5$ and the coherent component $b=0.5$.}
\end{figure}

\begin{equation}\label{E32}
\tilde{C}_{\omega\omega}(z)=|f(\omega)|^2\Big |\frac{\zeta_\omega(z)+\eta_\omega(z)}{z}\Big |.
\end{equation}
In consequence, the spectrum $W(\omega)$ is given by:
\begin{equation}\label{E33}
W(\omega)=|f(\omega)|^2|2Re\zeta_\omega(0)|.
\end{equation}
Sinse, in contradistinction to the case of the single Fano profile, the analytical formula for $W(\omega)$ for the model discussed here is extremely involved, we do not present it here. Instead, we consider the photoelectron spectra for two specific cases: an infinite value of the asymmetry parameter and a finite one. As in \cite{B2}, all the frequencies (energies) are given in units of $\Gamma$.\\

{\it Infinite value of the asymmetry parameter ($q\to\infty$)}.
For the degenerate case ($\omega_{21}=0$) the photoelectron spectrum is identical to that described in \cite{B3} for the case when the value of coherent component $b$ is fixed, while the value of the chaotic component $a_0$ is varied (figure 2). When the values of $a_0$ are kept constant and the values of $b$ are varied, the peaks are higher than in \cite{B3}. This is the case because equations (\ref{E14}a) and Eq. (\ref{E14}c) contain the coherent component $b$ which was omitted from the corresponding equations in \cite{B3} (see figure 3). When $a_0$ is small, i.e. the coherent part of the light dominates the fluctuations, the photoelectron spectra exhibit a characteristic Autler-Townes splitting.\\

Figures 4 and 5 present the photoelectron spectra for the nondegenerate case $(\omega_{21}\neq 0)$. The spectrum exhibits a two-peak structure regardless of the value of $b$. The right peak remains almost unchanged while the coherent and chaotic components are varied. The left peak increases with an increase in $b$ and decreases with an increase in $a_0$.\\
\begin{figure}[!h]
  \begin{center}
\includegraphics[scale=0.35]{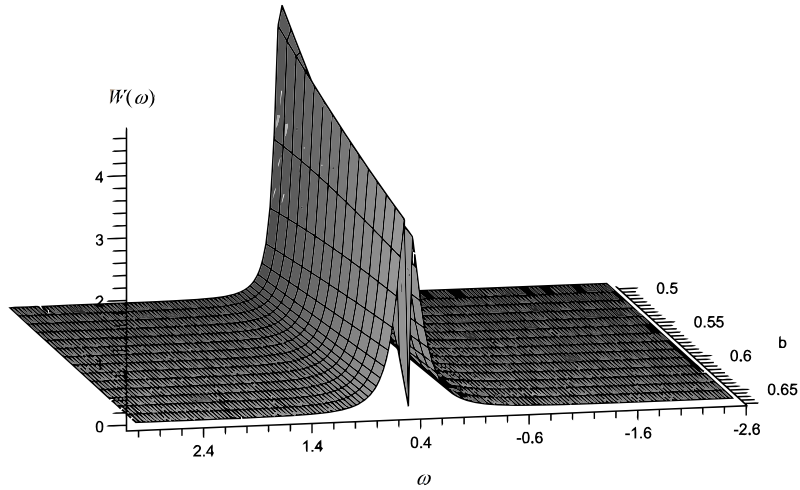}
\label{f6}
   \end{center}
  \caption{Photoelectron spectrum for the degenerate case ($\omega_1=\omega_2=0.5$) with the chaotic component $a_0=0$, $\omega_L=0.5$, autoionisation widths $\Gamma_1=\Gamma_2=0.5$ and asymmetry parameters $q_1=90$, $q_2=100$.}
\end{figure}
\begin{figure}[!h]
    \begin{center}
\includegraphics[scale=0.35]{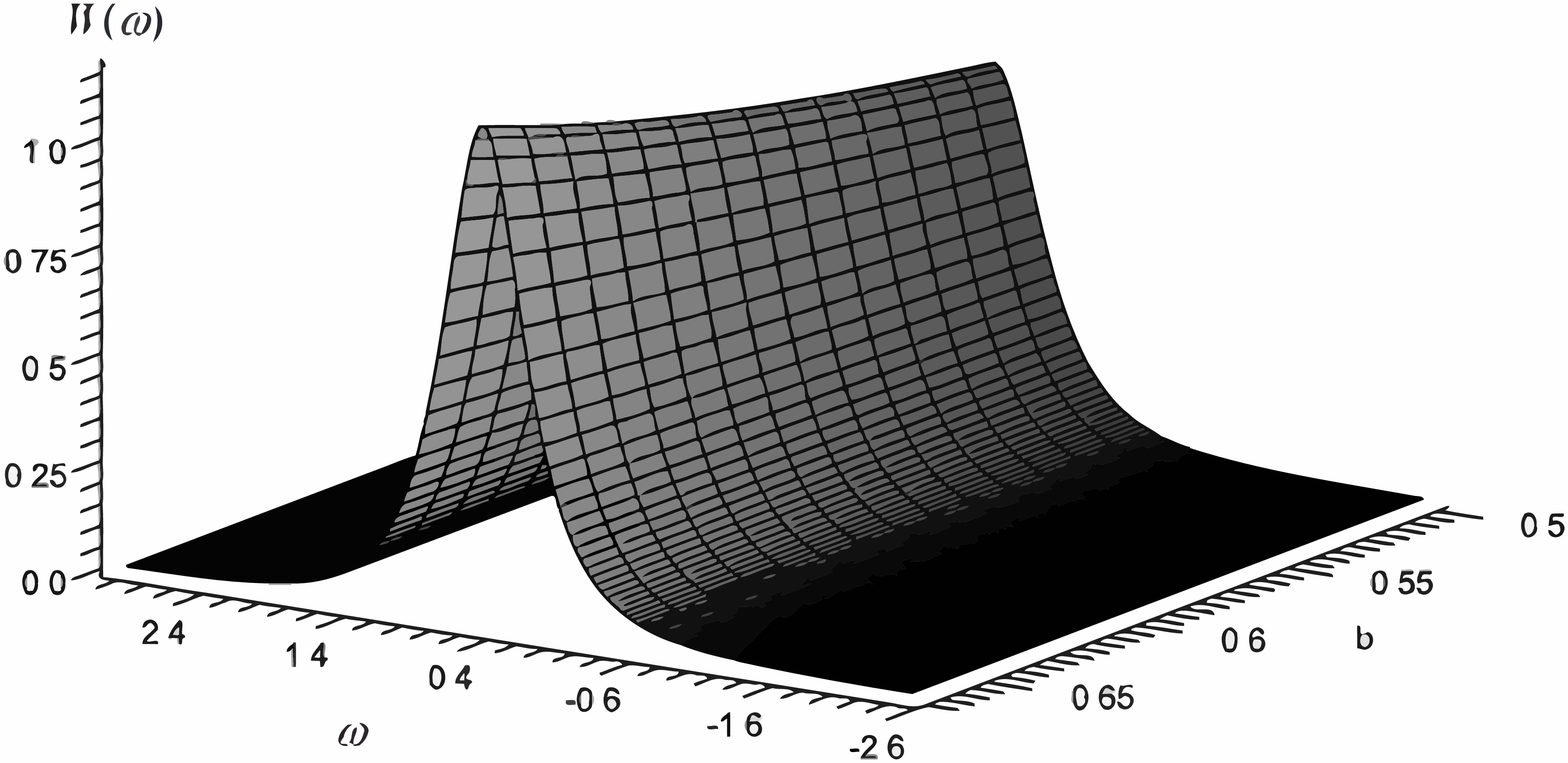}
\label{f7}
   \end{center}
  \caption{Photoelectron spectrum for the degenerate case ($\omega_1=\omega_2=0.5$) with the chaotic component $a_0=0.5$, $\omega_L=0.5$, autoionisation widths $\Gamma_1=\Gamma_2=0.5$ and asymmetry parameters $q_1=90$, $q_2=100$.}
\end{figure}
\begin{figure}[!h]
    \begin{center}
\includegraphics[scale=0.3]{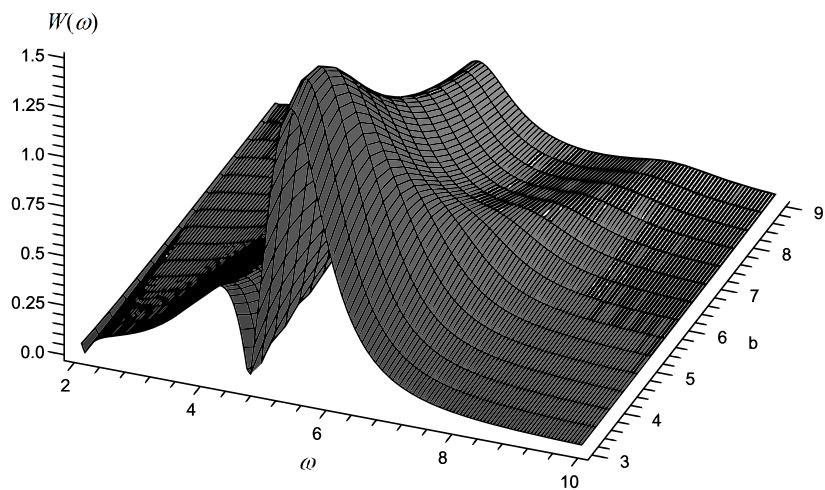}
\label{f8}
   \end{center}
  \caption{Photoelectron spectrum for $\omega_1=2.0$, $\omega_2=5.0$, $\omega_L=2.0$, autoionisation widths $\Gamma_1=0.1$, $\Gamma_2=10$ and the chaotic component $a_0=0.0$.}
\end{figure}
\begin{figure}[!h]
    \begin{center}
\includegraphics[scale=0.3]{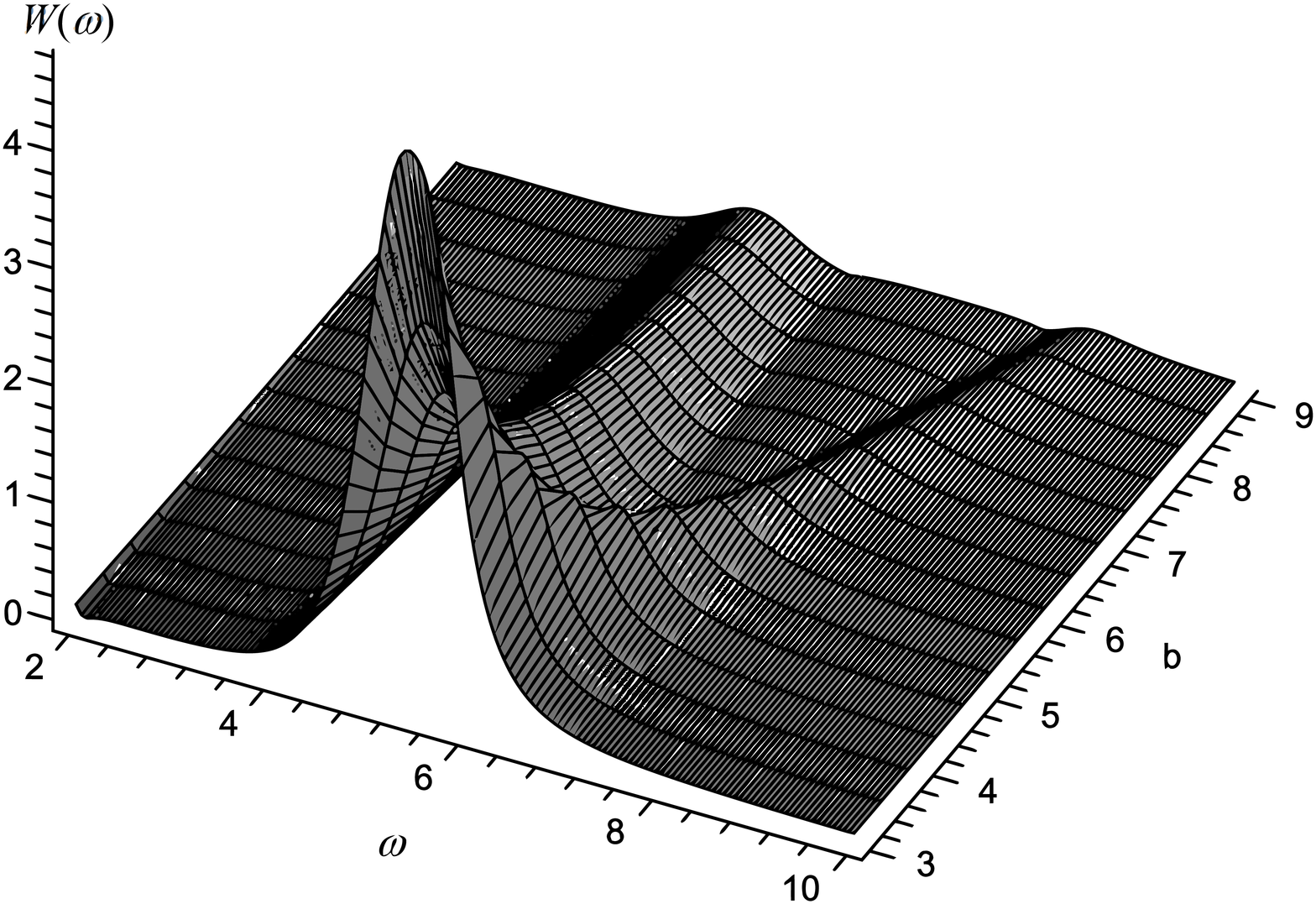}
\label{f9}
   \end{center}
  \caption{The same as in Fig.8 but for chaotic component $a_0=0.5$.}
\end{figure}
{\it Finite value of the asymmetry parameter}.
When $q_1$ and $q_2$ are finite, for the degenerate case $(\omega_{21}=0)$ the photoelectron spectra (figures 6 and 7) change considerably as we compare them with those discussed above. Figure 6 presents the photoelectron spectrum when the chaotic component is absent $(a_0=0)$ and for large values of $q$. For the weak field case, the symmetric spectrum is revealed and it consists of only one peak. When the field increases its amplitude, a sharp wedge in the spectrum appears. This spectrum exactly reproduces that obtained by Leonski et al \cite{B2}. However, when the chaotic component is present ($a_0\neq 0$) (figure 7), the sharp wedge in the spectrum disappears – only one, symmetric peak remains in the spectrum and its intensity is smaller than for the case $a_0=0$.\\

For the nondegenerate case, when the chaotic component is absent (figure 8) and for the smaller values of $b$ the spectrum exhibits two peaks. However, with an increase in $b$, the left peak quickly decreases rapidly, whereas the right peak increases. As the left peak disappears, the right peak begins to split into two peaks. When the chaotic component is present (figure 9), the two-peak structure of the spectrum becomes apparent with an increase in $b$. This process is faster than that discussed for the case when the chaotic component was absent. According to the Fano theory \cite{B1}, Fano zeros appearing in the photoelectron spectra manifest some quantum interferences which can occur in the system due to the fact that the system can achieve the continuum states via autoionization levels and by direct transitions. These interferences are weakened when the external electromagnetic field fluctuates. Therefore we can predict that in our model Fano zeros disappear in the photoelectron spectra. It follows from all presented figures that the Fano zeros really do not exist.\\ 
{\section{Conclusions}
In this paper, we considered a laser-induced autoionisation model, introduced earlier in \cite{B2}, where instead of one autoionising state, we consider two discrete states embedded in one continuum. This model can be referred to as the double Fano model. Since it is interesting to simulate quantum fluctuations by means of classical fluctuations, in this paper we assumeed that the laser light is decomposed into two parts: a deterministic (coherent) part and white noise, i.e. a randomly fluctuating chaotic component (cf \cite{B3}). Subsequently, we introduced and solved exactly a set of coupled stochastic integro-differential equations and thus obtained the results describing the dynamics of autoionization for the double Fano model. The results presented here contain certain additional terms, in contrast to those discussed in \cite{B3}, which make our findings more accurate. In particular, we derived exact analytical formulas for the photoelectron spectrum and compare the results with those discussed in [2, 3]. We considered the photoelectron spectra for an infinite value of the asymmetry parameter and, separately, for its finite value. Moreover, the considered spectra were discussed for the degenerate case $\omega_{12}=0$ and for the nondegenerate case ($\omega_{12}\neq 0$). We presented detailed result for both strong and weak fluctuations regimes. As in \cite{B15}, we believe that the model considered here is more realistic than that given in \cite{B2}, because the amplitudes of the real laser used in experiments always fluctuate.\\
\newpage{}


\begin{thebibliography}{99}
\bibitem{B1}
Fano U 1961~{\it Phys. Rev.}~{\bf
124}~1866
\bibitem{B2}
Leo\'nski W, Tana\'s R and Kielich S, 1987~{\it J. Opt. Soc.
Am.}~{\bf B 4}~72
\bibitem{B3}
Cao Long V and Trippenbach M 1986~{\it Z. Phys. B}~{\bf 63}~267
\bibitem{B4}
Rza\.zewski K and Eberly J H 1981~{\it Phys. Rev.
Lett.}~{\bf 47}~408\\
Rza\.zewski K and Eberly J H 1983~{\it Phys. Rev.}~{\bf A 27}~2026

\bibitem{B5}
Perina J Jr, Luks A, Leo\'nski W and Perinova V 2011~{\it Phys. Rev.}~{\bf A 83}~053416\\
Perina J Jr, Luks A, Leo\'nski W and Perinova V 2011~{\it Phys. Rev.}~{\bf A 83}~053430

\bibitem{B6}
Leo\'nski W, Tana\'s R and Kielich S, 1988~{\it J. Phys. B: At. Mol. Opt. Phys.}~{\bf 21}~2835

\bibitem{B7}
Leo\'nski W and Buzek V 1990~{\it J. Mod. Opt.}~{\bf 37}~1923

\bibitem{B8}
Leo\'nski 1993~{\it J. Opt. Soc. Am.}~{\bf B 10}~244

\bibitem{B9}
Raczy\'nski A, Rzepecka M, Zaremba J and Zielinska-Kaniasty S 2006 {\it Opt.Commun.}~{\bf 266} 552

\bibitem{B10}
Lewenstein M and Rza\.zewski K 2000 {\it Phys. Rev.}~{\bf A 61} 022105

\bibitem{B11}
Lu\'kyanchuk B {\it et al} 2010 {\it Nature Mater.}~{\bf 9} 707

\bibitem{B12}
Miroshnichenko A E, Flach S and Kivshar Y S 2010 {\it Rev. Mod. Phys.}~{\bf 82} 2257

\bibitem{B13}
W\'odkiewicz K 1985 {\it Noise in strong laser-atom interaction}~{\it Proc. 6th Int. School of Coherent Optics} (Ustro\'n, 19-26 Sept.)

\bibitem{B14}
Cao Long V and W\'odkiewicz K 1986 {\it J. Phys. B: At. Mol. Opt. Phys.}~{\bf 19} 1925

\bibitem{B15}
Doan Quoc K, Cao Long V and Leo\'nski W 2012 {\it Phys. Scr.}~{\bf T147} 014008

\end{thebibliography}
\end{document}